# Multiscale tunability of solitary wave dynamics in tensegrity metamaterials

Fernando Fraternali[*], Gerardo Carpentieri and Ada Amendola
*University of Salerno, Department of Civil Engineering*
*Via G. Paolo II, 132, 84084 Fisciano (SA), Italy.*

Robert E. Skelton[1] and Vitali F. Nesterenko[1,2]
[1]*University of California, San Diego, Mechanical and Aerospace Engineering Department*
[2]*University of California, San Diego, Materials Science and Engineering Program*
*9500 Gilman Dr., La Jolla, CA, USA 92093-0418.*

A new class of strongly nonlinear metamaterials based on tensegrity concepts is proposed and the solitary wave dynamics under impact loading is investigated. Such systems can be tuned into elastic hardening or elastic softening regimes by adjusting local and global prestress. In the softening regime these metamaterials are able to transform initially compression pulse into a solitary rarefaction wave followed by oscillatory tail with progressively decreasing amplitude. Interaction of a compression solitary pulse with an interface between elastically hardening and softening materials having correspondingly low-high acoustic impedances demonstrates anomalous behavior: a train of reflected compression solitary waves in the low impedance material; and a transmitted solitary rarefaction wave with oscillatory tail in high impedance material. The interaction of a rarefaction solitary wave with an interface between elastically softening and elastically hardening materials with high-low impedances also demonstrates anomalous behavior: a reflected solitary rarefaction wave with oscillatory tail in the high impedance branch; and a delayed train of transmitted compression solitary pulses in the low impedance branch. These anomalous impact transformation properties may allow for the design of ultimate impact mitigation devices without relying on energy dissipation.

PACS numbers: 05.45.Yv, 43.25.+y, 45.50.-j, 46.40.Cd

## I. INTRODUCTION

A novel area of research has emerged over the last few years regarding the design and manufacturing of structural lattices modulated with periodic elastic moduli and mass densities. It has been shown that such linear elastic metamaterials may exhibit anomalous acoustic behaviors, like negative effective elastic moduli; negative effective mass density; acoustic negative refraction; phononic band gaps; and local resonance, to name just a few examples ([1] and the references therein).

The dynamics of strongly nonlinear metamaterials with power-law interaction law between elements has also been investigated ([2]-[10]). Elastically hardening (or stiffening) discrete systems with exponent $n$ greater than one ("normal" materials) support compressive solitary waves and unusual reflection of wave on material interfaces. While elastically softening systems with $n < 1$ ("abnormal" materials) support the propagation of rarefaction solitary waves under initially compressive impact loading ([2],[11]). Small-scale cellular composite materials featuring elastic softening and extremely large ratio of elastic modulus to density have been recently assembled using an additive manufacturing approach [12].

Ordinary engineering materials typically exhibit either elastic hardening (e.g., crystalline solids), or elastic softening (e.g., foams). More versatile is the geometrically nonlinear response of structural lattices based on tensegrity units (e.g., tensegrity prisms), which may gradually change their elastic response from hardening to softening through modification of mechanical, geometrical and prestress variables ([13][15]).

In this Letter, we numerically investigate the dynamics of periodic lattices of lumped masses connected by tensegrity prisms exhibiting either softening or hardening elastic response tuned by local and global prestress ([14],[15]). We show that such systems are able to support tunable solitary rarefaction and compression waves exhibiting anomalous wave transmission and reflection from interfaces between branches with different acoustic impedances. The observed behaviors pave the way to the optimal design of tunable tensegrity metamaterials, and ultimate impact protection devices that do not require energy dissipation.

## II. TUNABILITY OF TENSEGRITY UNITS

Let us consider a few millimeter scale tensegrity prism (or "tensegrity unit"), which is composed of three titanium alloy Ti6Al4V struts (or bars), and nine PowerPro® Spectra fibers (or strings). Each base of the prism is in frictionless contact with a metalling disc of thickness $d$ acting as a lumped mass ($m$) (FIG. 1). We examine prisms using 0.28 mm Spectra fibers, and 0.8 mm circular bars, which can be manufactured through electron beam melting [16]. Let us assume that the tensegrity unit is uniformly loaded in compression by axial forces with their resultant $F$ applied in the center of mass of the terminal bases. Under such loading, the deformation of the unit maintains its top and bottom bases parallel to each other and changes the angle of twist $\varphi$ and the height $h$. Effective tensegrity placements with all strings in tension (or under zero force) correspond to the angle of twist interval $\varphi \in [\frac{2}{3}\pi, \pi]$, with the bars getting in touch with each other for $\varphi = \pi$ ("locking configuration") [13][14]. The configuration corresponding to zero external force instead features an angle of twist $\varphi = \varphi_0 = \frac{5}{6}\pi$, and its internal prestress can be tuned through the cross-string prestrain $p_0$ ("local prestrain" determined by the pretensioning of the cross-strings in the assembling phase) [14][15]. Experimental tests have shown that the post-locking behavior of tensegrity prisms may lead to a plateau regime with axial deformation increasing under almost zero axial force increments [15].

[*]Electronic address: f.fraternali@unisa.it

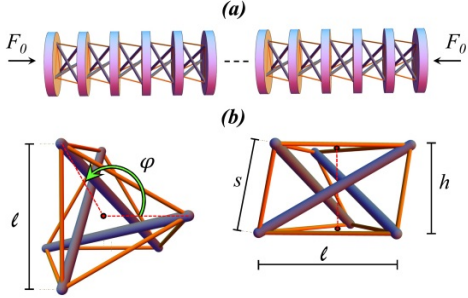

FIG. 1 (color online). (a) Chain of tensegrity prisms (strongly nonlinear springs) and lumped masses globally prestressed by force $F_0$. (b) Top view (on the left) and side view (on the right) of the tensegrity unit. Local forces within the unit are shown in Supplemental Material.

Let $\varepsilon$ denote the strain of the tensegrity unit (or "unit strain") defined as $\varepsilon = (h_0 - h)/h_0$, where $h_0$ denotes the prism height for $F = 0$. Curves in FIG. 2 illustrate the axial force vs unit strain curves corresponding to different values of the local prestrain $p_0$ obtained following the quasi-static approach presented in [14] (see Supplemental Material for the adopted geometrical and material properties). For $p_0 \leq 0.02$, the quasi-static response of the tensegrity unit features a hardening branch near the origin $\varepsilon=0$ ($dF/d\varepsilon$ increasing with increasing $\varepsilon$). Such a branch is followed by a softening regime ($k_\varepsilon$ decreasing with increasing $\varepsilon$), for larger values of the unit strain. The unit response is instead always softening for $p_0 \geq 0.03$. The effective modulus of the unit goes to zero for $\varepsilon \to 0$ and $p_0 = 0$. Similar behavior ($dF/d\varepsilon = 0$) corresponds to the turning points of the force-strain curves. In all such cases, the metamaterial in FIG. 1 behaves as a "sonic vacuum" [2].

It is instructive to employ power-laws of the form $F = C_n \cdot \varepsilon^n$ to approximate data sets extracted from the tensegrity unit response in correspondence with different local and global prestrains. For $p_0 = 0$ and $\varepsilon \in [0.05, 0.15]$, we observe that such a response is best fitted by a hardening power-law with $n = 2.25$ and $C_n = 0.61$ kN. The response for $p_0 = 0.04$ and $\varepsilon \in [0.15, 0.25]$ is instead best fitted by a softening power-law with $n = 0.74$ and $C_n = 0.11$ kN. In both cases, the unit exhibits positive tangent stiffness ($dF/d\varepsilon > 0$), i.e. statically stable behavior. For $p_0 = 0.06$ and $\varepsilon \in [0.45, 0.55]$ the response of the tensegrity unit turns to unstable ($dF/d\varepsilon < 0$), and is best fitted by a power-law with negative exponent ($n = -0.31$, $C_n = 0.04$ kN).

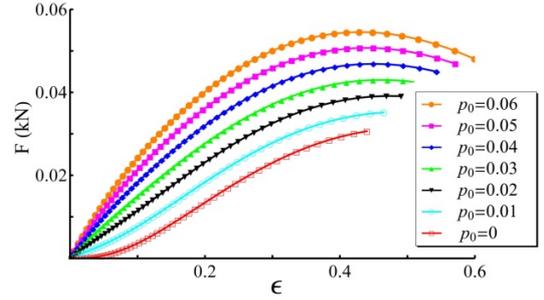

FIG. 2 (color online) Multiscale tunability of the quasi-static force response of the tensegrity unit to unit strain $\varepsilon$ at different values of the local prestrain $p_0$.

### III. ANOMALOUS DYNAMICS OF TENSEGRITY LATTICES

A periodic lattice composed of tensegrity units and lumped masses subjected to a static precompression force $F_0$ was dynamically excited. An impact velocity $v_0$ was imposed to the mass center of the first disc in order to reproduce the effects of an impulsive compressive loading. This initial condition is also corresponding to the delta force function applied to the mass center of the first disc. The assumption of frictionless contact between the units and the lumped masses implies that no residual torques or bending moments are transmitted from the units to the masses [14], which therefore may be assumed to move only in the longitudinal direction.

Hereafter, we use the symbol $m_1$ representing the combined mass of a tensegrity prism ($m_0$) and a disc ($m$). On assuming the mass ratio $m_1/m_0 = 300$, we modeled the lattice in FIG. 1 as a chain of point masses connected by massless spring. The latter feature the mechanical response of the tensegrity unit in compression, and zero response in tension. We characterized the deformation of the structure through the "system strain" $\varepsilon_s = (H_0 - H)/H_0$, where $H = h + d$; $H_0 = h_0 + d$. We let $\varepsilon_{s_0}$ denote the value of $\varepsilon_s$ induced by $F_0$ ("global prestrain"). By keeping $\varepsilon_{s_0}$ equal to 0.01, we found that the ratio between the effective elastic modulus and the effective density of the system in FIG. 1 ranges from 0.85 m$^2$/s$^2$ (effective modulus: $E = 2.64$ kPa; effective density: $\rho = 3.09$ g/cm$^3$) to 295 m$^2$/s$^2$ ($E = 767.13$ kPa; $\rho = 2.60$ g/cm$^3$), as $p_0$ increases from 0 to 0.25. Much larger elastic moduli might be featured by tensegrity prisms endowed with rigid bases ([14]-[15]). The dynamic behavior of the tensegrity unit was approximated by its quasistatic response due to the significant difference of characteristic time of waves duration and characteristic period of the tensegrity unit ensured by the value of lumped masses (see data in Supplemental Material).

We first investigated the wave dynamics of lattices showing 1400 tensegrity units exhibiting elastic-softening response. The strain pulses generated in chains featuring local prestrain $p_0 = 0.04$ and global prestrain $\varepsilon_{s_0} = 0.15$ are shown in FIG. 3 for different impact velocities. The specific acoustic impedance $\rho c_0$ of the current chain is equal to 20.89 kPa.s/m, $c_0$ denoting the speed of sound.

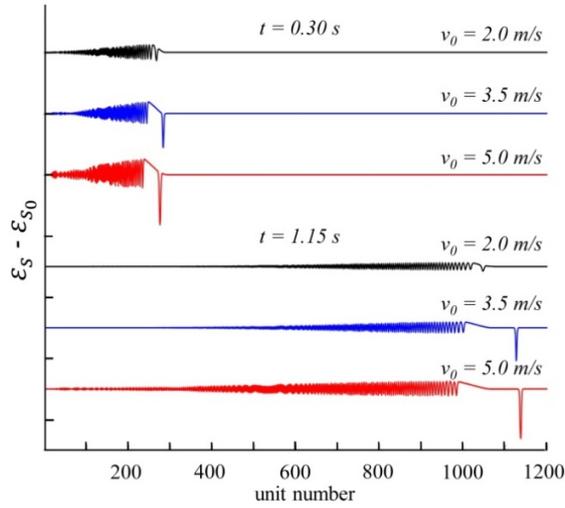

FIG. 3 (color online). Elastic-softening chain. Evolution of initial compression pulse (not shown) into rarefaction wave and periodic train at different impact velocities (global and local prestrain are correspondingly equal to $\varepsilon_{s_0} = 0.15$, $p_0 = 0.04$). The strain is offset for visual clarity ($Y$ ticks indicate 0.2).

For both impact velocities $v_0 = 5.0$ m/s and $v_0 = 3.5$ m/s, we observe the formation of a leading rarefaction soliton of relevant amplitude followed by a dispersive, oscillatory tail led by a nonstationary compression wave [2]. A similar wave dynamics is observed also in the case with $v_0 = 2.0$ m/s, but in such a case the rarefaction soliton and the oscillatory tail have smaller amplitudes and leading rarefaction solitary wave did not separate from the compression wave at investigated distances of their propagation. The rarefaction soliton moves faster than the oscillatory tale and compression wave and separates from them within about 300 units under impact with velocities 5 and 3.5 m/s. For $v_0 = 5.0$ m/s we notice that the minimum strain accomodated by the lattice is equal to $-0.014$ at $t = 1.15$ s, which implies that locally the system assumes negative strains, still being globally compressed. But at $v_0 = 3.5$ m/s and $v_0 = 2.0$ m/s the system instead remains both locally and globally compressed. The total width of the leading rarefaction pulse spans 10 units in the cases with $v_0 = 5.0$ m/s and $v_0 = 3.5$ m/s, and 12 units in the case with $v_0 = 2.0$ m/s, assuming a cutoff of $|\varepsilon_s - \varepsilon_{s_0}|/\varepsilon_{s_0} = 0.98$ (see Supplemental material for additional data). As shown in [11], the size of a solitary rarefaction soliton amounts to 7 and 11 units in power law materials showing $n = 0.50$ and $n = 0.80$, respectively. Referring to the case with $v_0 = 5.0$ m/s, we observe that the average speed of the rarefaction soliton is supersonic (1.11 $c_0$), while the average speed of the compression pulse immediately following the rarefaction soliton is slightly subsonic (0.97 $c_0$). The results in FIG. 3 suggest that the impacts at 5.0 m/s and 3.5 m/s generate strongly nonlinear rarefaction waves with strain amplitude approximatively equal or larger than the global prestrain in the system. The oscillatory tail in all cases is getting longer and its amplitude approaches zero as the propagation distance increases. The process of compression pulse transformation can be accelerated by increasing the local prestrain $p_0$ (see Supplemental Material).

We now examine the behavior of a compression solitary pulse approaching the interface between two chains in tensionless contact with each other: a Low Impedance chain of 700 masses connected by tensegrity units with Elastic Hardening response (LIEH branch: $\varepsilon_{s_0} = 0.01$; $p_0 = 0$; $\rho c_0 = 2.86$ kPa.s/m), and a High Impedance chain of 700 masses connected by tensegrity units with Elastic Softening response (HIES branch: $\varepsilon_{s_0} = 0.15$; $p_0 = 0.05$; $\rho c_0 = 21.22$ kPa.s/m). FIG. 4(a) shows the evolution of strain pulses in the examined system under the impact velocity $v_0 = 5.0$ m/s. We observe that initial solitary compression wave travels along the LIEH branch. The interaction of such waves with the LIEH-HIES interface generates a gap between the branches and unexpected transmitted solitary rarefaction waves with oscillatory tail in the HIES branch. A train of reflected solitary compression waves is observed in the LIEH branch (see Supplemental Material for numerical data). The reflection of the compression solitary wave back as a compression solitary wave was expected, due to the difference in the acoustic impedance between the two branches [2],[3],[6]. What was unexpected is that the solitary compression wave is reflected as a train of solitary compression waves. Such an anomalous reflection is new and different from that observed in the interaction of compression solitary waves with material interfaces in strongly nonlinear granular media [3],[6].

Our final results deal with the interaction of a rarefaction solitary wave with an interface between a HIES lattice of 5000 masses and LIEH lattice of 1000 masses under the impact with velocity $v_0 = 5.0$ m/s. FIG. 4(b) shows that HIES system supports an incident solitary rarefaction pulse with oscillatory tail. Its interaction with the given interface results in a train of transmitted solitary compression waves in LIEH system. A reflected solitary rarefaction wave followed by oscillatory tail propagates back into the HIES branch. The results in FIGs. (3) and 4(b) suggest that the oscillatory tail of the incident rarefaction pulse might degenerate into an infinitely small amplitude oscillatory tail.

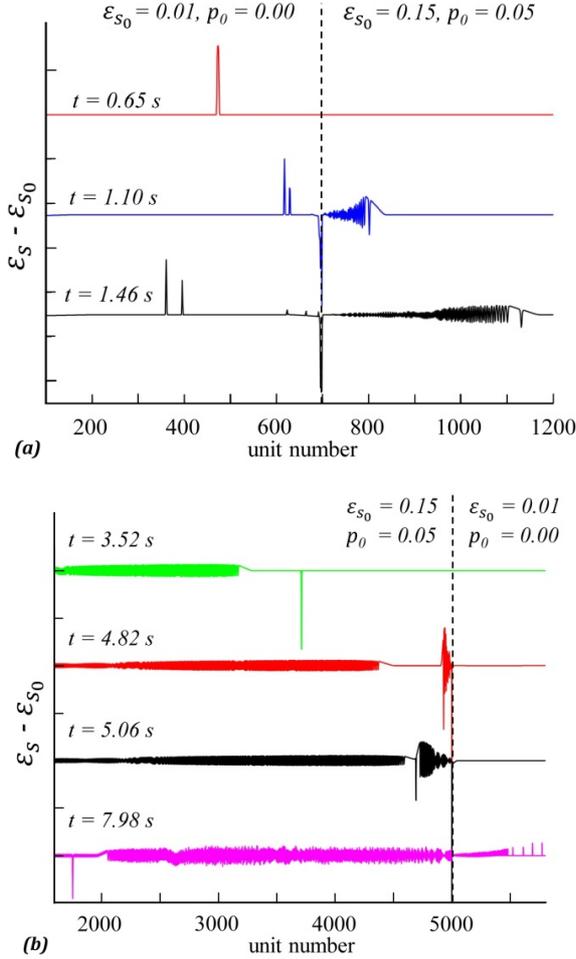

FIG. 4 (color online). (a) Interaction of a compression solitary wave with a LIEH-HIES interface ($Y$ ticks indicate 0.2); (b) Interaction of a rarefaction solitary wave with a HIES-LIEH interface ($Y$ ticks indicate 0.1). The strain is offset for visual clarity.

## IV. CONCLUSIONS

We numerically investigated the wave dynamics of novel multiscale tunable metamaterials which feature tensegrity prisms playing the role of nonlinear springs connecting lumped masses. We showed that such systems provide physical realizations of either normal and abnormal power-law materials [2],[11], being able to dramatically change their geometrically nonlinear response from elastically hardening to elastically softening [14],[15]. The presented results highlight that softening tensegrity metamaterials may transform an initially compressive disturbance into a rarefaction wave of finite amplitude with progressively vanishing oscillatory tail. We demonstrated anomalous reflection of compression and rarefaction solitary waves from interfaces of two tensegrity based metamaterials. The analyzed systems are ultimate impact mitigation systems which do not require dissipation of energy, but a relatively large number of units (of the order of 1000 in the examined cases), as a function of local and global prestress. If the size of the units can be scaled down to about 10 μm (via, e.g., laser litography [17]) we expect that an effective impact protection barrier would require a total length of 10 mm. The wave dynamics presented here pave the way to future research focusing on the design of strongly nonlinear metamaterials based on tensegrity concepts [13],[15].

We thank the Italian Ministry of Foreign Affairs for financial support (Grant No. 00173/2014).

# Supplementary Data:
# Multiscale tunability of solitary wave dynamics in tensegrity metamaterials

Fernando Fraternali[1*], Gerardo Carpentieri[1], Ada Amendola[1], Robert E. Skelton[2], Vitali F. Nesterenko[2,3]

[1]*University of Salerno, Department of Civil Engineering*
*Via G. Paolo II, 132, 84084 Fisciano (SA), Italy*
[2]*University of California, San Diego, Mechanical and Aerospace Engineering Department*
*9500 Gilman Dr. MC-0411, La Jolla, CA, USA 92093-0411*
[3]*University of California, San Diego, Materials Science and Engineering Program*
*9500 Gilman Dr., La Jolla, CA, USA 92093-0418*

The present supplement includes four tables (Tables S1-S4); four figures (Figures S1-S4); and three movies (movie_Fig3, movie_Fig4a and movie_Fig4b, see section Animations of www.fernandofraternaliresearch.com).

Table S1 provides the geometrical and mechanical properties of the system represented in Fig. 1, while Tables S2-S4 illustrate numerical data associated with Figs. 3 and 4 of the main paper.

Fig. S1 illustrates the deformation history and the internal forces in a tensegrity unit for a given value of the internal prestrain $p_0$. Fig. S2 illustrates the effects of the internal prestrain tuning on the wave dynamics of tensegrity lattices featuring elastic-softening response. Fig. S3 shows the dynamics of a variant of the system analyzed in Fig. 4a of the main paper, consisting of a LIEH branch with 700 masses and a HIES branch with 2200 masses. Fig. S4 illustrates the dynamics of a variant of the system analyzed in Fig. 4b, which includes a HIES branch with 1500 masses and a LIEH branch with 2000 masses.

The supplemental movies show animations of Figs. 3, 4a and 4b of the main paper.

**Notation**

$E_b$: Young modulus of the bars

$E_s$: Young modulus of the strings

$s_N$: rest length of the cross-strings

$\ell_N$: rest length of the base-strings

$b_N$: rest length of the bars

$R$: radius of the terminal discs

$d$: thickness of the terminal discs

$m_0$ mass of a single tensegrity prism (not including the terminal discs)

$m$ mass of a disc (lead, density=11340 kg/m$^3$)

$H_0$: height of the unit cell (prism + lumped mass) under zero external force ($F = 0$)

$H_{\varepsilon_0}$: height of the unit cell under the precompression force $F_0$

$k_{\varepsilon_0} = \left.\frac{dF}{dH}\right|_{H=H_{\varepsilon_0}}$: stiffness at the precompressed state

$c_0 = H_{\varepsilon_0}\sqrt{k_{\varepsilon_0}/m_1}$: speed of sound of the system at the precompressed state ($m_1 = m + m_0$)

$T_w = H_{\varepsilon_0}/c_w$: characteristic time of waves duration ($c_w$: wave speed)

$T_0 = 2\pi\sqrt{m_0/k_{\varepsilon_0}}$: oscillation period of the tensegrity unit

**Table S1:** Geometrical and mechanical properties of the system in FIG. 1 of the main paper.

| $s_N$ [mm] | $\ell_N$ [mm] | $b_N$ [mm] | $R$ [mm] | $d$ [mm] | $E_b$ [GPa] | $E_s$ [GPa] | $m_0$ [g] | $m$ [g] |
|---|---|---|---|---|---|---|---|---|
| 6.00 | 8.70 | 11.50 | 18.66 | 2.00 | 120.00 | 5.48 | 0.08 | 24.82 |
| $p_0$ | 0.00 | 0.02 | 0.04 | 0.06 | 0.08 | 0.10 | 0.12 | 0.14 |
| $H_{\varepsilon_0}$ [mm] | 7.41 | 7.52 | 7.63 | 7.74 | 7.86 | 7.97 | 8.08 | 8.19 |

**Table S2:** Properties of the solitary waves in FIG. 3 of the main paper
($H_{\varepsilon_0}$ = 7.63 mm; $k_{\varepsilon_0}$ = 20.973 kN/m; $c_0$ = 5.97 m/s).

| $v_0$(m/s) | type | amplitude ($\varepsilon_s - \varepsilon_{s_0}$) | width (units) | wave speed / $c_0$ | $T_w/T_0$ |
|---|---|---|---|---|---|
| 2.0 | $(R)^{(1)}$ | 0.114/0.122[3] | 12[4] | 1.00[5] | 2.75[5] |
|  | $(C)^{(2)}$ | 0.171/0.157[3] | - | 0.99[5] | 2.80[5] |
| 3.5 | $(R)^{(1)}$ | 0.037/0.040[3] | 10[4] | 1.08[5] | 2.55[5] |
|  | $(C)^{(2)}$ | 0.186/0.166[3] | - | 0.98[5] | 2.82[5] |
| 5.0 | $(R)^{(1)}$ | -0.016/-0.014[3] | 10[4] | 1.11[5] | 2.48[5] |
|  | $(C)^{(2)}$ | 0.198/0.170[3] | - | 0.97[5] | 2.86[5] |

(1) solitary rarefaction wave
(2) compressive pulse leading the oscillatory tail
(3) values at $t = 0.30$ s / $t = 1.15$ s
(4) average value for $t \in [0.90, 1.15]$ s
(5) average value for $t \in [0.30, 1.15]$ s

**Table S3:** Properties of the solitary waves in FIG. 4(a) of the main paper
(LIEH branch: $H_{\varepsilon_0}$ = 7.40 mm; $k_{\varepsilon_0}$ = 0.394 kN/m; $c_0$ = 0.93 m/s;
HIES branch: $H_{\varepsilon_0}$ = 7.69 mm: $k_{\varepsilon_0}$ = 21.652 kN/m; $c_0$ = 6.11 m/s).

| $v_0$(m/s) | type | amplitude ($\varepsilon_s - \varepsilon_{s_0}$) | width (units) | wave speed / $c_0$ |
|---|---|---|---|---|
|  | $(I)^{(1)}$ | 0.31[4] | 7[4] | 5.796[6] |
| 5.0 | $(R)^{(2)}$ | 0.23[5] | 4[5] | 5.717[7] |
|  | $(T)^{(3)}$ | -0.07[5] | 7[5] | 0.986[7] |

(1) incident solitary compression wave
(2) leading reflected solitary compression wave
(3) transmitted solitary rarefaction wave
(4) value at $t = 0.6$ s
(5) value at $t = 1.4$ s
(6) average value for $t \in [0.5, 0.6]$ s
(7) average value for $t \in [1.1, 1.3]$ s

**Table S4:** Properties of the solitary waves in FIG. 4(b) of the main paper.
(HIES branch: $H_{\varepsilon_0}$ = 7.68 mm: $k_{\varepsilon_0}$ = 21.652 kN/m; $c_0$ = 6.11 m/s;
LIEH branch: $h_{\varepsilon_0}$ = 7.35 mm; $k_{\varepsilon_0}$ = 0.395 kN/m; $c_0$ = 0.93 m/s).

| $v_0$(m/s) | type | amplitude ($\varepsilon_s - \varepsilon_{s_0}$) | width (units) | wave speed / $c_0$ |
|---|---|---|---|---|
|  | $(I)^{(1)}$ | −0.17[4] | 10[4] | 1.147[6] |
| 5.0 | $(R)^{(2)}$ | −0.08[5] | 5[5] | 1.083[7] |
|  | $(T)^{(3)}$ | 0.07[5] | 3[5] | 3.891[7] |

(1) leading incident solitary rarefaction wave
(2) reflected solitary rarefaction wave
(3) leading transmitted solitary compression wave
(4) average value for $t \in [0.90, 1.15]$ s
(5) value at $t = 1.3$ s
(6) average value for $t \in [0.5, 0.6]$ s
(7) average value for $t \in [1.1, 1.3]$ s

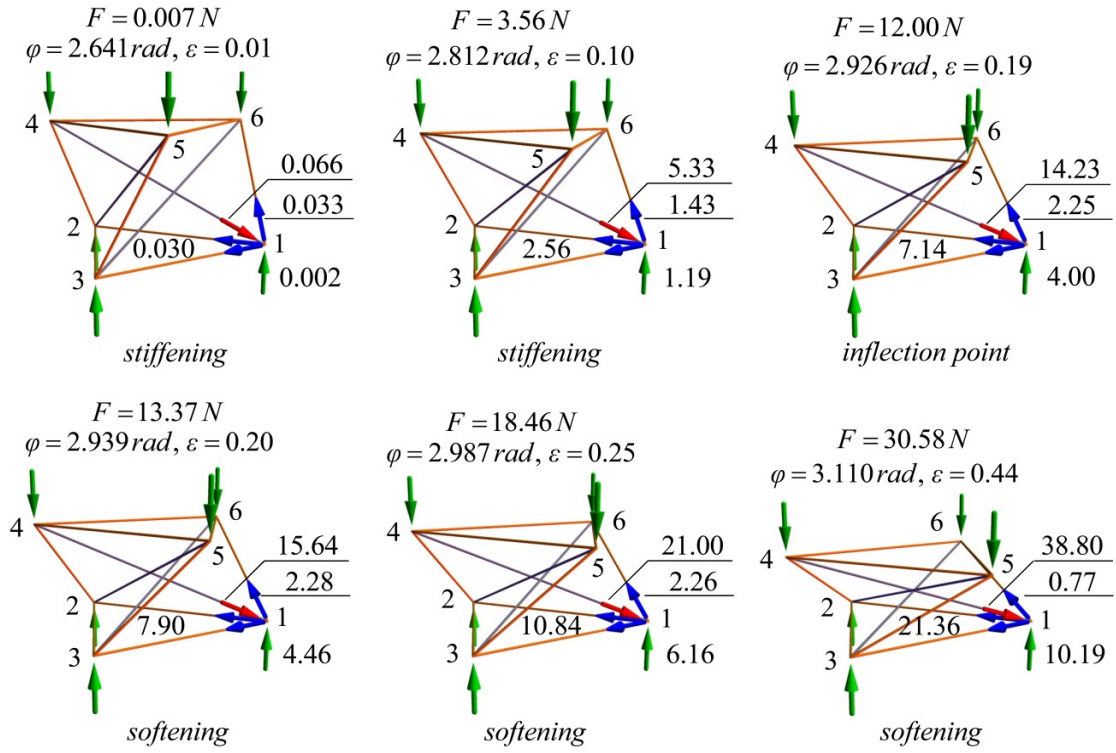

**FIG. S1.** Sequence of configurations of the tensegrity unit under compressive loading ($p_0 = 0$), and corresponding internal forces acting on the individual members (N).

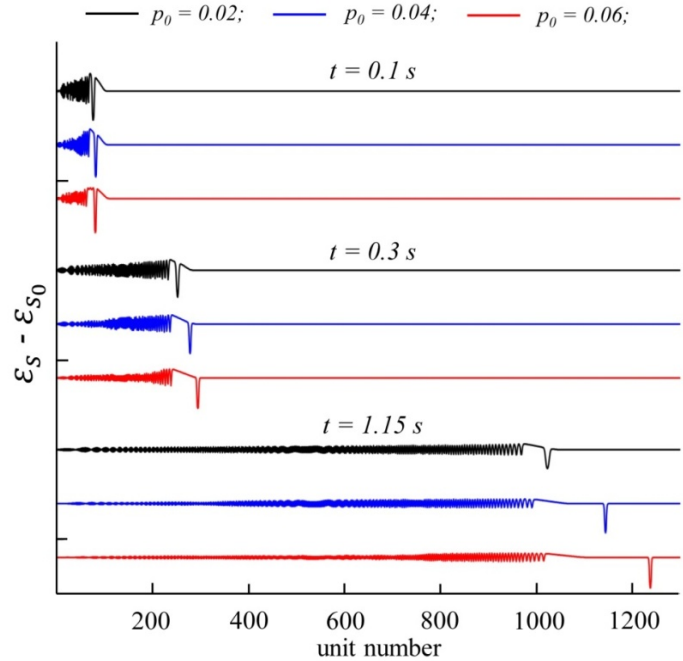

**FIG. S2.** Tuning of internal prestress $p_0$ for fixed global prestrain $\varepsilon_{s_0} = 0.15$ and $v_0 = 5.0$ m/s. The strain is offset for visual clarity (Y ticks indicate 0.5).

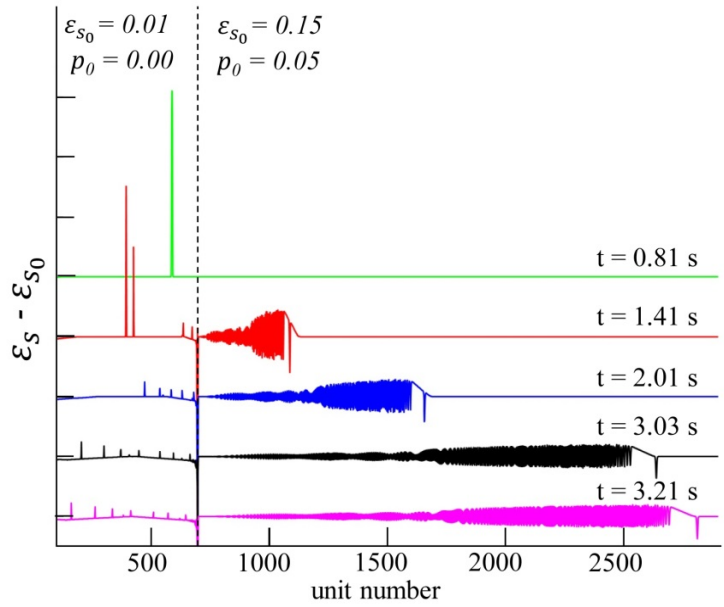

**FIG. S3**. Interaction of a rarefaction solitary wave with the interface between a LIEH branch with 700 masses and a HIES branch with 2200 masses (Y ticks indicate 0.1).

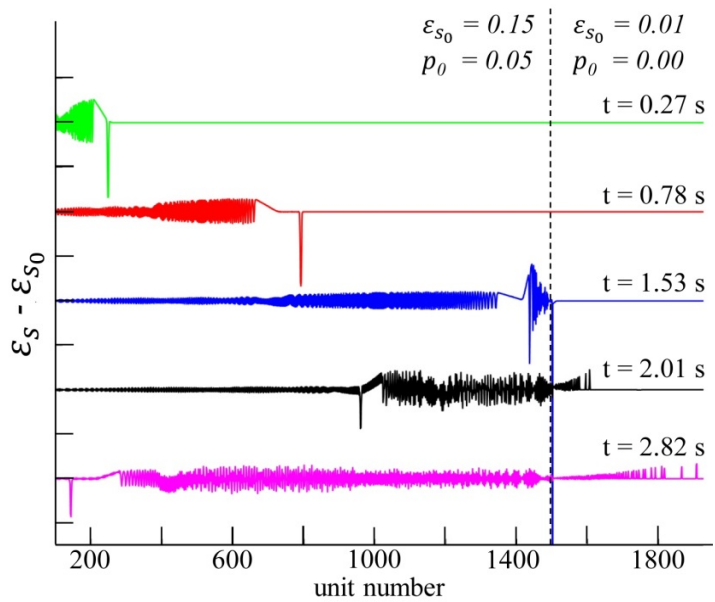

**FIG. S4.** Interaction of a rarefaction solitary wave with the interface between a HIES branch with 1500 masses and a LIEH branch with 2000 masses (Y ticks indicate 0.1).